# Metal-organic kagome systems as candidates to study spin liquids, spin ice or the quantum anomalous Hall effect

*Adam Hassan Denawi[1, 2,3] *, Xavier Bouju[3], Mathieu Abel [1], Johannes Richter [4,5], and Roland Hayn[1]*

[1]Aix Marseille Université, CNRS, IM2NP UMR 7334, 13397 Marseille, France.

[2]CEA Paris-Saclay, Service de Recherches de Métallurgie Physique, 91191 Gif-sur-Yvette, France

[3] Centre d'élaboration de matériaux et d'études structurales (CEMES), CNRS, Université de Toulouse, Toulouse, France

[4] Institut für Physik, Universität Magdeburg, P.O. Box 4120, D-39016 Magdeburg, Germany

[5] Max-Planck Institut für Physik Komplexer Systeme, Nöthnitzer Str. 38, D-01187 Dresden, Germany

**ABSTRACT**

We present the results of first-principle calculations using the Vienna Ab-initio Simulation Package (VASP) for a new class of organometallics labeled $TM_3C_6O_6$ (TM =Sc, Ti, V, Cr, Fe, Co, Ni and Cu) in the form of planar, two-dimensional, periodic free-standing layers. These materials, which can be produced by on-surface coordination on metallic surfaces, have a kagome lattice of TM ions. Calculating the structural properties, we show that all considered materials have local magnetic moments in the ground state, but four of them (with Fe, Co, Ni and Cu) show spin-crossover behavior or switch between magnetic and nonmagnetic states by changing the lattice constant, which could be valuable for possible epitaxy routes on various substrates. Surprisingly, we find a very large richness of electronic and magnetic properties, qualifying these materials as highly promising metal-organic topological quantum materials.

We find semi-conductors with nearest-neighbor ferromagnetic (FM) or antiferromagnetic (AFM) couplings for V, and Sc, Ti and Cr, respectively, being of potential interest to study spin ice or spin liquids on the 2D kagome lattice. Other TM ion systems combine AFM couplings with metallic behavior (Fe and Ni) or are ferromagnetic kagome metals like $Cu_3C_6O_6$ with band crossings at the Fermi surface. For the latter compound, the spin orbit coupling is shown to be responsible for small gaps which makes them a candidate material to observe the quantum anomalous Hall effect (QAHE).

**INTRODUCTION**

The kagome Heisenberg antiferromagnet is a well-known paradigm of highly frustrated magnetism [1–5]. Furthermore, the electronic and transport properties on the kagome lattices attracted an enormous interest recently, since they are characterized by symmetry protected band crossing points similar to the Dirac point in graphene [6]. So, the kagome systems are promising representations for topological quantum materials that became particularly clear with the recent discovery of ferromagnetic kagome metals [7]. Together with spin-orbit coupling one expects topological [8] and Chern insulating phases, which are interesting for the quantum anomalous Hall effect (QAHE) [9]. These exotic states can be traced back to several peculiarities of the kagome lattice consisting of corner sharing triangles, namely frustration, and the possibilities for Dirac points and of local excitations of magnetic or electronic nature leading to dispersionless (flat) excitation branches. Such flat-band systems have attracted much attention in several areas of physics, and many interesting phenomena that are related to flat bands have been found, see, e.g., the reviews [10–13]. Interestingly, for partial filling of a nearly dispersionless band one may expect fractional topological quantum states [14,15] in analogy to the fractional quantum Hall effect. It is remarkable that also the antiferromagnetic state in the kagome compounds $Mn_3Ge$ or $Mn_3Sn$ allows the anomalous Hall effect [16–19].

Whereas numerous topological quantum materials were already found and investigated in the class of inorganic layered crystals, there is much less progress for organometallic lattices despite existing proposals [20]. However, the kagome systems $Mn_3C_6O_6$ and $Cu_3C_6O_6$, as well as $Fe_3C_6O_6$ have recently been synthesized by on-surface coordination reaction on noble metal surfaces [21-23]. These metal-organic coordination networks consist of $C_6O_6$ rings and a dense kagome lattice of 3d transition metal (TM) ions. In the present theoretical study, we extend the search of exotic systems to all 3d TM ions going from Sc to Cu. We study here free-standing monolayers to serve as reference systems for the adsorbed case. It is promising that we found good candidates in the material class of $TM_3C_6O_6$ for realizations of several unconventional kagome systems.

As already mentioned, the kagome Heisenberg antiferromagnet is highly frustrated. In the classical limit its ground state is massively degenerated, i.e. it has an extensive ground-state manifold [24]. Quantum fluctuations may select coplanar ground states [25,26]. In the extreme quantum limit (spin quantum numbers $S=1/2$ and $S=1$) the ground state is magnetically disordered. Although there is a plethora of theoretical studies for $S=1/2$, see, e.g. Refs. [27,28] and references therein, the nature of the quantum ground state is still under debate. Favored candidates are a gapless U(1) Dirac spin liquid [29,30] and a gapped Z2 spin liquid [31,32]. For $S=1$ there are much less studies available, however, the absence of magnetic order seems to be well established, see e.g. [33-36] and references therein. Candidates for the ground states without magnetic long-range order are a chiral spin liquid, a hexagonal-singlet solid or a trimerized state [34-36]. For larger spin $S>1$ there are indications for a magnetically-ordered ground state [33,37,38].

In addition to the extensively discussed nature of the spin-liquid ground state, the intriguing magnetization process of the kagome Heisenberg antiferromagnet has attracted much attention. For $S=1/2$ magnetization exhibits plateaus at 1/3, 5/9 and 7/9 of the saturation magnetization [39,40] and a macroscopic jump at the saturation field due to the very existence of a flat one-magnon band [41]. The 1/3-plateau as well as the jump at the saturation are also present for $S>1/2$ but both shrink with increasing $S$ [41,42]. A spectacular feature of the quantum kagome Heisenberg antiferromagnet is the appearance of a magnon crystal phase just below the saturation field [43,44].

The above outlined theoretical predictions for kagome Heisenberg antiferromagnet at zero and finite magnetic fields go hand in hand with numerous experimental studies, see Refs. [4,45] for an overview. Among many candidate materials Herbertsmithite is a near-perfect $S=1/2$ kagome Heisenberg antiferromagnet compound showing characteristic features of a spin liquid [2,46]. There are also experimental indications for the plateaus and the magnon crystallization [47,48] predicted by theory. Turning to FM nearest neighbor interactions, they are interesting as well. Classical spins on the kagome lattice with a FM nearest neighbor interaction and site-dependent single-site anisotropy have a residual entropy at zero temperature and, therefore, they are called spin ice [49].

In the present study we investigate free-standing $TM_3C_6O_6$ monolayers by ab-initio band structure calculations taking into account the Coulomb interaction in the 3d shell and clarify its structural, electronic and magnetic properties. Most importantly, we found candidates for realizations of very different exotic kagome systems: the semiconducting Sc system with AFM interactions between spins 1/2 allowing the spin liquid state, several AFM kagome systems with different values of the local spin going from $S=1/2$ to $S=5/2$ with the exception of $S=3/2$ (V system) for which the interaction is FM and could give rise to spin ice, and finally $Cu_3C_6O_6$ which is predicted to be a FM kagome metal and is a candidate for the QAHE.

We also find spin-crossover (SCO) complexes and transitions between magnetic and nonmagnetic states in the studied material class. These SCO complexes [50–57] have gathered considerable attention for their potential use as intrinsic switch to build nanoscale electronic components. Indeed, in our case, they are composed of a central transition-metal (TM) ion surrounded by $C_6O_6$ units whose spin state can be potentially switched by applying external stimuli such as temperature, light, pressure, magnetic or electric fields, or current [58–69]. In this paper, we theoretically explore the synergy between the two research fields (SCO and topological quantum materials) by studying the spin properties of the $TM_3C_6O_6$ structures. The results uncover $TM_3C_6O_6$ as a fascinating material class with a surprisingly rich behavior mentioned above. That is outlined below based on our calculations of (i) the total energies for different lattice constants and spins states, (ii) the various densities of states, (iii) the total and local magnetic moments and the magnetic couplings, and (iv) the band structures without and with spin-orbit (SO) coupling.

**METHOD**

We used the Vienna Ab-initio Simulation Package (VASP) [70] at the level of the spin-polarized generalized-gradient approximation (SGGA) in the form of the Perdew, Burke, and Ernzerhof (PBE) functional. The SGGA functional is used to investigate the structural properties (lattice constants, atomic positions, spin-crossover transitions). It is well known, however, that the standard Density Functional Theory (DFT) has difficulties to describe correctly the electronic density of states, including the gap values, especially for 3d transition metal ions. For that reason, we apply here also the Spin polarized Generalized-Gradient Approximation with a Hubbard term $U$ (SGGA+$U$). The SGGA+$U$ corrections were introduced by Liechtenstein et al [71]. We take in the following $U$ = 5 eV and an exchange energy of $J$ = 0.90 eV. The necessity of the SGGA + $U$ method for metal−organic compounds with transition metal ions is proven by many examples such as TM-TCNB [72,73] or TM-TCNQ [74] TM-ZQ [75–79] where TM is a transition metal, and the organic molecules are tetracyanobenzene (TCNB), tetracyanoquinone (TCNQ) and zwitterionic quinone (ZQ).

The interaction between the valence electrons and ionic cores was described within the framework of the projector augmented wave (PAW) method [80–83]. The electronic wave functions were expanded in plane waves with a kinetic energy cutoff of 480 eV, the monolayers were relaxed and the convergence criteria for the energy deviations was $10^{-7}$ eV. The Gaussian smearing method was used in these calculations and a width of $\sigma$ =0.01 eV was adopted in most calculations. The Brillouin zone was usually determined by a set of 6×6×1 k-points in the unit cell using the Monkhorst-Pack points [84]. The geometry optimization was performed without spin-orbit coupling. But a relativistic calculation was used for $Cu_3C_6O_6$ to investigate the possibility of a QAHE state. That high-precision band-structure calculation was performed with 18×18×1 k-points and SIGMA=0.001 eV.

## STRUCTURAL PROPERTIES

The first step in our calculations is to perform the structural optimization of the $TM_3C_6O_6$ networks studied by calculating the total energy as a function of lattice constant $a$ and to determine the fundamental state. For that purpose, we use the SGGA functional, but the influence of the Hubbard $U$ correction on the structural properties is only small. The iteration process is repeated until the calculation of the total energy converges. Figure 1 shows the total energies for the ferromagnetic (FM) state and the non-magnetic (NM) configurations for those situations where we found a SCO situation, i.e. for Fe and Ni, or a transition between magnetic and nonmagnetic states like for Co and Cu. Since the Hubbard $U$ term has no influence for a nonmagnetic state, we have to use the SGGA method (without $U$) to compare magnetic and nonmagnetic states in Fig. 1.

The crystal structure of $TM_3C_6O_6$ is shown in Fig. 2(a). The grey, red, and green spheres represent carbon, oxygen, and transition metal atoms, respectively. The lattice constant is optimized and the resulting lattice constants are presented in Table 1 including the $U$ correction, and in the appendix A without $U$. Also, the metal-oxygen distances are listed there and one can see that they remain nearly constant throughout the 3d series. We find two compounds, namely $Co_3C_6O_6$ and $Cu_3C_6O_6$, with two minima, a nonmagnetic state at the equilibrium lattice constant of 7.45 Å (8.25 Å) for the Co (Cu)-system, and a magnetic state with $S = 3/2$ per Co ($S = 1/2$ per Cu) by variation of the lattice constant. The magnetic state has an energy minimum for a lattice constant of 7.74 Å (7.78 Å). There are transitions between states of different spin for Fe between $S = 2$ and $S = 1$ and for Ni between $S = 1$ and $S = 0$, which one can qualify as SCO transitions due to the competition between crystal field and Coulomb energies. Notice that the crystal structure of nonmagnetic $Cu_3C_6O_6$ at the second minima with a lattice constant of 8.25 Å is slightly different from Fig. 2(a). The $C_6O_6$ rings are rotated such that two Cu-O bonds are stronger than the two others [22].

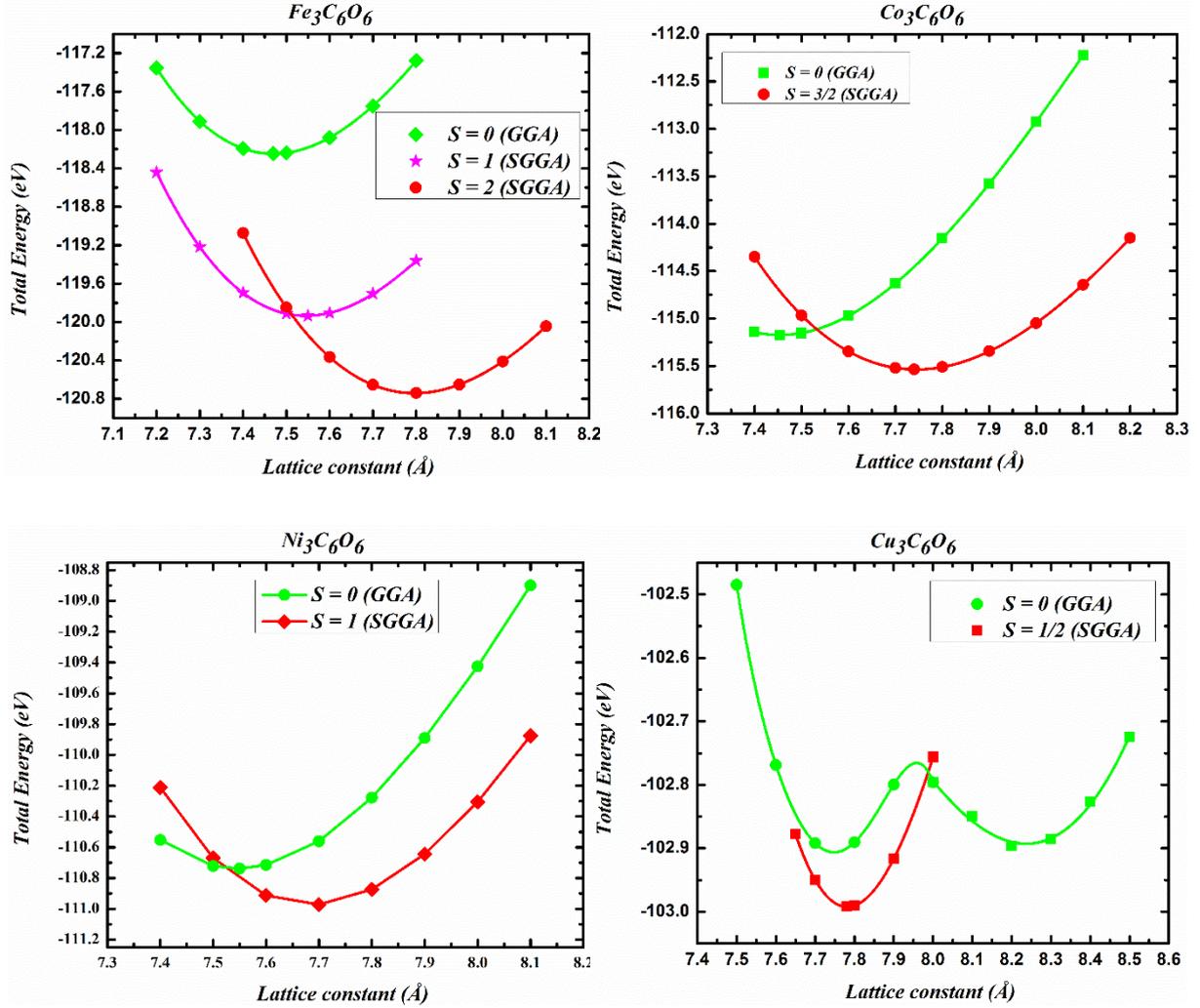

*Figure 1:* Total energy versus lattice constant for the free-standing $TM_3C_6O_6$ structures calculated with the SGGA.

## MAGNETIC PROPERTIES

In Fig. 2 we show the charge density that varies between 0 and 0.5 Å$^{-3}$ for the cut plane in the x-y plane. The spin density varies between -0.02 and 0.02 Å$^{-3}$ for the cut plane (see Fig. 3). The charge density is defined as the sum of the electron densities for spin up and spin down ($\rho_{\text{spin-up}}$ + $\rho_{\text{spin-down}}$). The magnetism distributions of the benzene hexa-carboxylic ($C_6O_6$) molecules with TM metals can be intuitively studied by the analysis of the spin density, which is defined as the difference between the electron densities for spin up and spin down ($\rho_{spin-up}$ - $\rho_{spin-down}$). The change in the magnetization distribution when one goes from Sc to Cu is quite interesting. In the beginning of the 3d series (Sc, Ti, V, and Cr) the spin density is nearly exclusively concentrated on the metal sites, which generates a kagome lattice. For Fe, Co, and Ni, some moments appear at the oxygen sites that bridge the metals. And finally, for $Cu_3C_6O_6$, the main magnetic moment is located on the carbon ring.

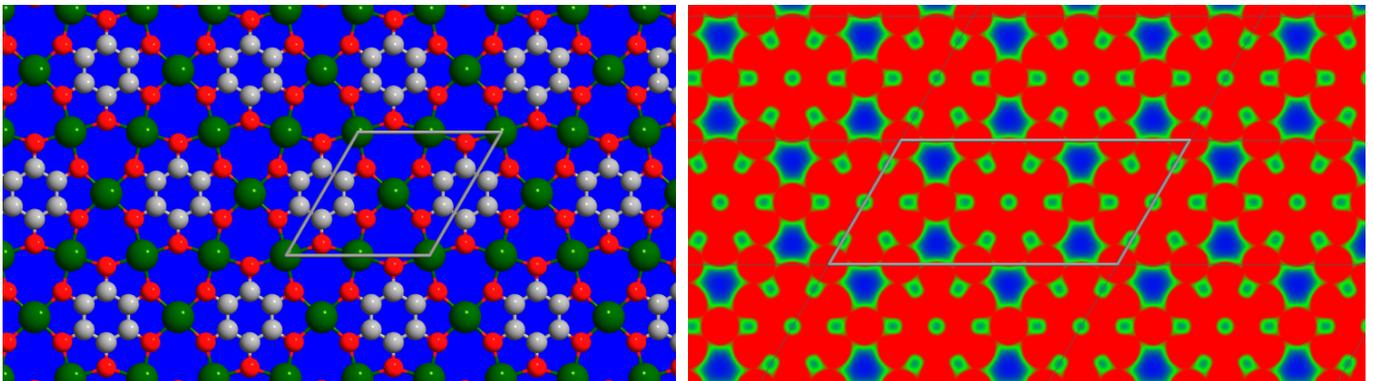

***Figure 2:*** *Geometrical structure and charge density of a free-standing structure of TM atoms and $C_6O_6$ molecule (C: grey, O: red, and TM: Green).*

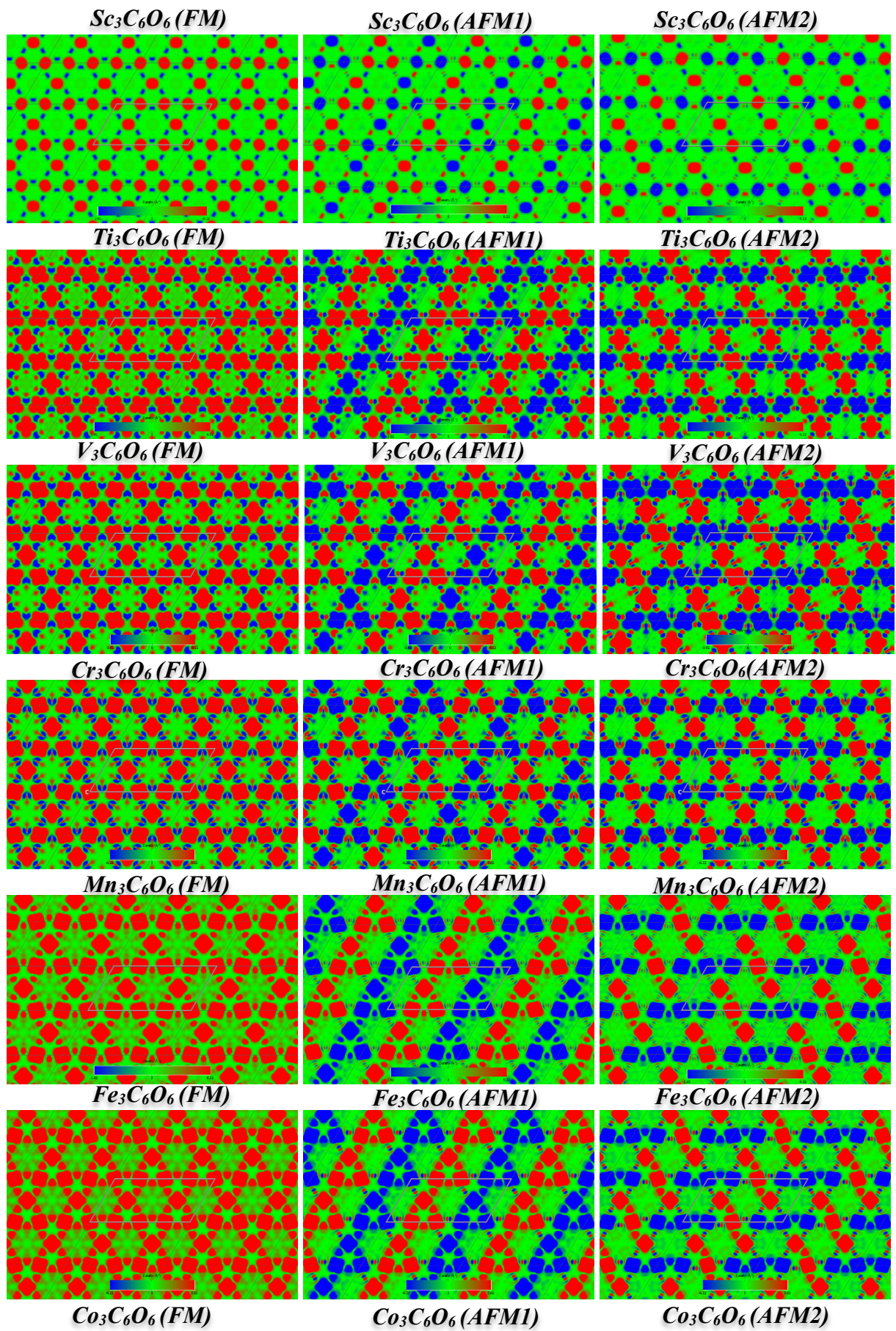

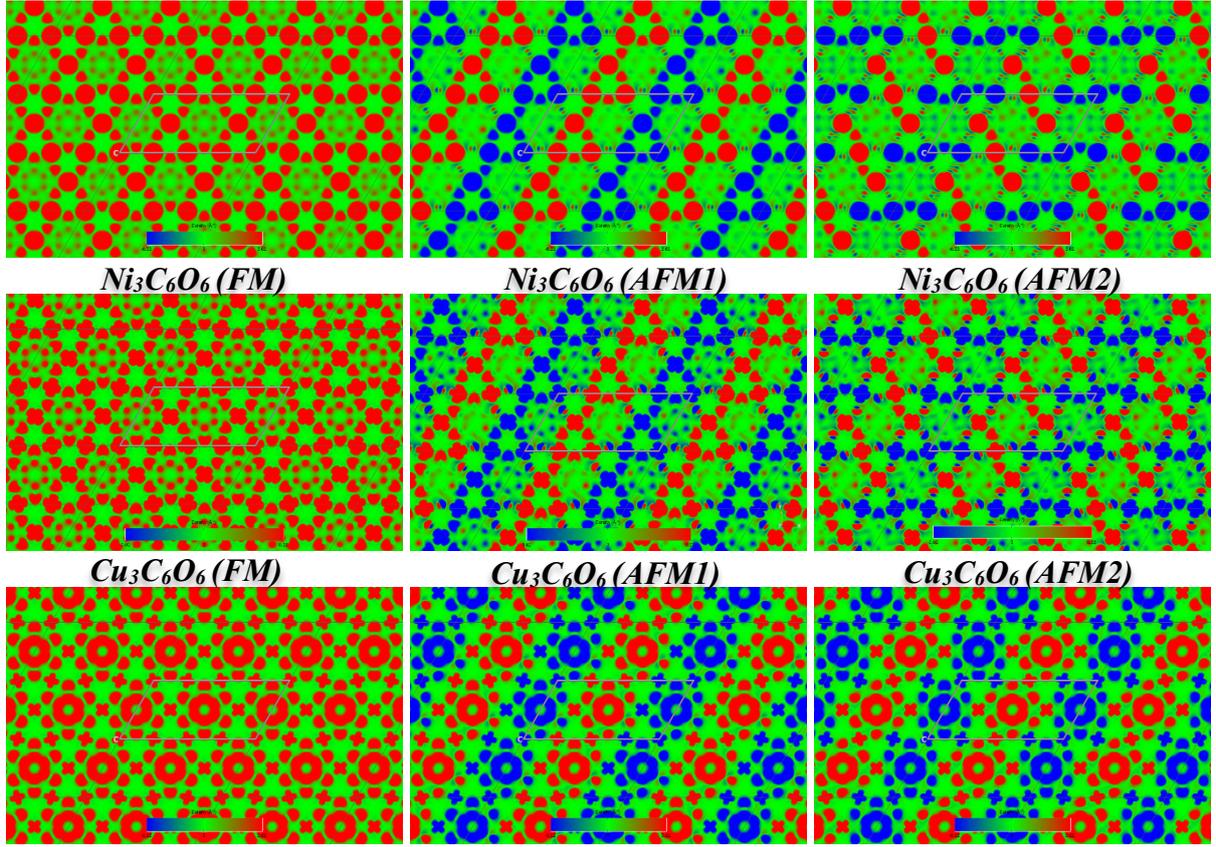

*Figure 3: Cut plane of the spin density contribution with isovalue of ±0.02 of TM$_3$C$_6$O$_6$. The spin surface is shown in the x-y plane like the atomic structure. The ferromagnetic arrangement and two different antiferromagnetic arrangements are shown and the results are obtained with SGGA+U.*

To describe the magnetic moments at the TM sites, we calculated the exchange couplings $J_1$ and $J_2$ to nearest and second nearest neighbors on the kagome lattice (see Fig. 4a) in the Heisenberg Hamiltonian

$$H = \sum_{\langle i,j \rangle} J_{ij}\vec{S}_i\vec{S}_j = \frac{1}{2}\sum_{i,j} J_{ij}\vec{S}_i\vec{S}_j = \frac{1}{2}\sum_{r,g} J_g\vec{S}_r\vec{S}_{r+g}$$

where $\langle i,j \rangle$ means that one sums over each bond only once. For that purpose, we estimate the energy difference between the ferromagnetic state and the antiferromagnetic one after relaxation of all atomic positions $E_{ex} = E_{AF} - E_{FM}$, defined per magnetic ion. If that energy difference is positive (negative), it indicates that the ferromagnetic (antiferromagnetic) state is preferred. For a ferromagnetic arrangement of the spins $S$ on the kagome lattice, we obtain $E_{FM} = 2(J_1 + J_2)S^2$ whereas the two antiferromagnetic arrangements that are visible in Fig. 3 give $E_{AF1} = \frac{2}{3}(J_1 - J_2)S^2$ and $E_{AF2} = -\frac{2}{3}(J_1 + J_2)S^2$. Taking the two energy differences $E_{ex,1} = E_{AF1} - E_{FM}$ and $E_{ex,2} = E_{AF2} - E_{FM}$, we calculate the corresponding exchange couplings reported in Table 1.

With the SGGA+$U$ method, the exchange couplings are antiferromagnetic and of short range for the transition metals Sc, Ti, Cr, and Fe. We find a rather large value of $J_1$=48 meV for Sc, whereas $J_1$ varies between 0.66 meV (Cr) and 3.66 meV (Ti) for the other three compounds. In all cases, the second neighbor coupling is at least a factor of ten smaller. We find it remarkable that the SGGA+$U$ values of exchange couplings are confirmed by the SGGA calculations (see Appendix A). In two cases, we find the energy of the ferromagnetic configuration lower than both antiferromagnetic solutions, for the V- and the Cu-system. Furthermore, the second neighbor coupling considerably exceeds the first neighbor one for $Cu_3C_6O_6$ and we cannot exclude further reaching exchange couplings which means that a short-range Heisenberg model is probably not applicable in that situation. But the metallic state and the strong energy gain of the FM configuration are strong indications for a FM ground state of $Cu_3C_6O_6$. For the $Co_3C_6O_6$ structure, the ferromagnetic state is energetically located in between both antiferromagnetic ones ($E_{ex,1}$ = -11.93 meV and $E_{ex,2}$ = 21.31 meV). It means that the exchange coupling to nearest neighbors $J_1$ = -11.08 meV is ferromagnetic, but the second neighbor exchange coupling $J_2$ = 7.53 meV turns to an antiferromagnetic exchange (see Table 1).

*Table 1. Distance between the TM atoms and the O atoms ($d_{TM-O}$, in Å), lattice constant (a, in Å), energy differences $E_{ex,1/2}=E_{AFM\,(1/2)}-E_{FM}$ per TM atom, in meV, exchange coupling constants $J_1$ and $J_2$ as explained in the text in meV, total magnetic moments (M per TM atom, in $\mu_B$), local magnetic moments of the d orbital at the TM atoms ($M_d$ per TM atom, in $\mu_B$), local magnetic moments at the TM atoms ($M_m$ per TM atom, in $\mu_B$), energy band gaps (spin up ($E_a$) and spin down ($E_b$), in eV), total energy gaps ($E_g$, in eV), for 2D $TM_3C_6O_6$ with the PAW-SGGA+U method. The exchange constants $J_1$ and $J_2$ are put in parentheses for $Cu_3C_6O_6$ since a Heisenberg model description is questionable there.*

| U=5eV J=0.9 eV | $d^1$ $Sc_3C_6O_6$ (S=1/2) | $d^2$ $Ti_3C_6O_6$ (S=1) | $d^3$ $V_3C_6O_6$ (S=3/2) | $d^4$ $Cr_3C_6O_6$ (S=2) | $d^5$ $Mn_3C_6O_6$ (S=5/2) | $d^6$ $Fe_3C_6O_6$ (S=2) | $d^7$ $Co_3C_6O_6$ (S=3/2) | $d^8$ $Ni_3C_6O_6$ (S=1) | $d^9$ $Cu_3C_6O_6$ (S=1/2) |
|---|---|---|---|---|---|---|---|---|---|
| $d_{TM-O}$ | 2.14 | 2.09 | 2.08 | 2.05 | 2.10 | 2.04 | 2.03 | 2.03 | 2.03 |
| a (Å) | 8.06 | 7.92 | 7.90 | 7.82 | 7.98 | 7.82 | 7.76 | 7.72 | 7.75 |
| $E_{ex,1}$ | -17.91 | -5.36 | 2.17 | -4.02 | -6.08 | -15.60 | -11.93 | -31.32 | 24.86 |
| $E_{ex,2}$ | -33.91 | -10.24 | 5.45 | -7.55 | -10.87 | -28.92 | 21.31 | -50.53 | 30.50 |
| $J_1$ | 48.00 | 3.66 | -1.09 | 0.66 | 0.57 | 2.50 | -11.08 | 14.41 | (16.92) |
| $J_2$ | 2.87 | 0.18 | 0.19 | 0.05 | 0.08 | 0.21 | 7.53 | 4.54 | (-62.67) |
| M | 1 | 2 | 3 | 4 | 5 | 4 | 3 | 2 | 1 |
| $M_d$ | 0.415 | 1.428 | 2.36 | 3.37 | 4.58 | 3.61 | 2.66 | 1.67 | 0.397 |
| $M_m$ | 0.454 | 1.495 | 2.39 | 3.41 | 4.65 | 3.65 | 2.68 | 1.69 | 0.394 |
| $E_a$ | 1.15 | 2.07 | 1.60 | 1.71 | 1.22 | 1.84 | 1.68 | 0 | 2.23 |
| $E_b$ | 2.25 | 2.20 | 2.22 | 2.22 | 1.50 | 0 | 1.13 | 0 | 0 |
| $E_g$ | 0.43 | 1.42 | 1.60 | 1.67 | 1.22 | 0 | 1.13 | 0 | 0 |

## ELECTRONIC PROPERTIES

The spin-resolved band structure calculations for the $TM_3C_6O_6$ structures were carried out in the high symmetry directions of the first Brillouin zone and they are shown together with the total densities of states (DOS) in Fig. 5. Figure 6 shows the partial DOS. We present the results with the SGGA+*U* functional which we think to be more relevant than the SGGA results as outlined above. The gap values in Fig. 5 are also given in Tab. 1. It is important to note that the kagome lattice described here (Fig. 4a) has the same point group symmetry $D_{6h}$ as graphene. As a consequence, the numerous band crossings at the K point are guaranteed by symmetry, which qualifies all materials of the $TM_3C_6O_6$ class to be potentially interesting topological quantum materials. Note also, that the TM-O lattice in Fig. 2a coincides exactly with the Cu-O lattice in the planes of Herbertsmithite [46], a naturally occurring material that is widely studied to observe spin liquid behavior.

The local magnetic moment is mostly determined by the filling of the 3d levels which is also important for the electronic properties. The splitting of those 3d levels is best visible in a GGA calculation (nonmagnetic solution without $U$ correction) which is free of exchange or Coulomb shifts (Fig. 4b). Following the analysis in Ref. [9] we can distinguish the $d_{z^2}$ band, a doubly degenerate $d_{xz}/d_{yz}$ band and the $d_{x^2-y^2}/d_{xy}$ complex which splits into a lower lying bonding and a higher lying antibonding part due to the metal-metal interaction in the plane. The order of these levels is rather constant throughout the 3d series, just the $d_{z^2}$ band moves a little bit from one system to another.

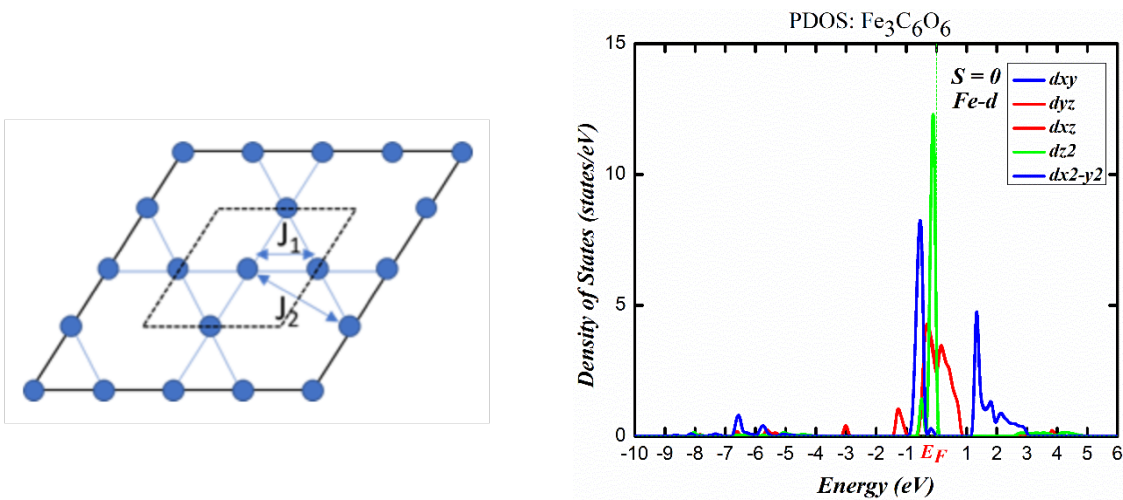

***Figure 4:*** *Schematic representation of the kagome lattice with the exchange couplings $J_1$ and $J_2$ (a, left) and partial, orbital resolved 3d DOS for $Fe_3C_6O_6$ in GGA (no spin-polarized and without U correction) to illustrate the ligand field splitting (b, right).*

As one can easily observe in Figs. 5 and 6, the Fe, Ni and Cu structures are predicted to be metallic. Therefore, they could be interesting for technological applications as metal-organic conducting layers. One might wonder, why $Fe_3C_6O_6$ is metallic whereas its neighboring compounds in the 3d series with Mn and Co are both insulating. Furthermore, the gap of the Fe case is not very sensitive to the Hubbard $U$ parameter in the 3d shell. That is explained since the relevant band at the Fermi level has very few 3d contribution, both for the Fe and the Co cases, but that band is half-filled for $Fe_3C_6O_6$ and completely occupied for $Co_3C_6O_6$.

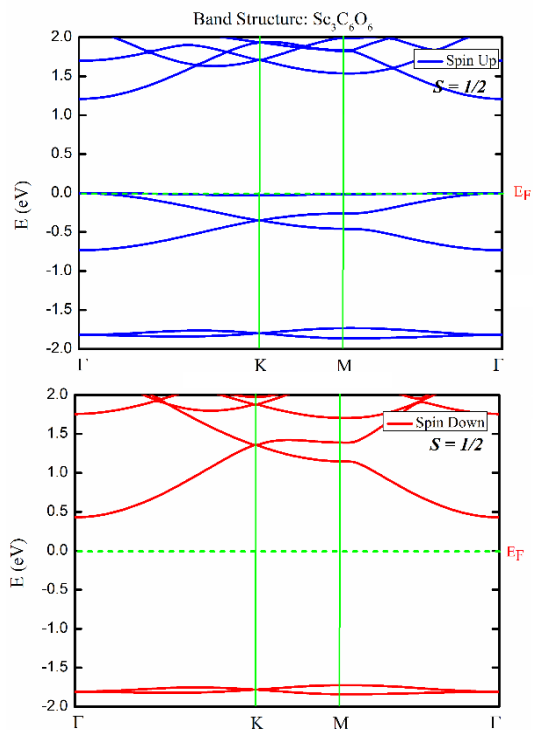
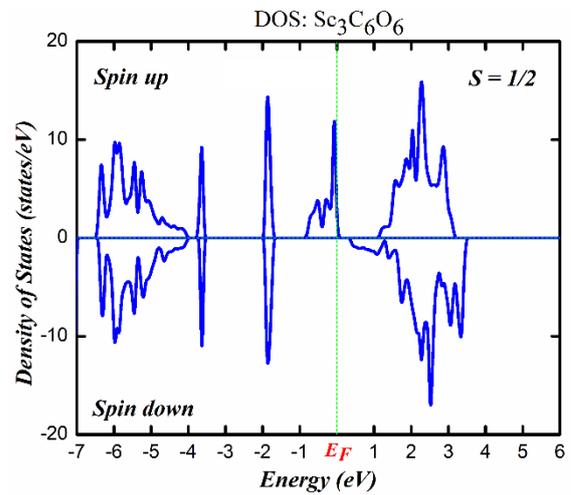

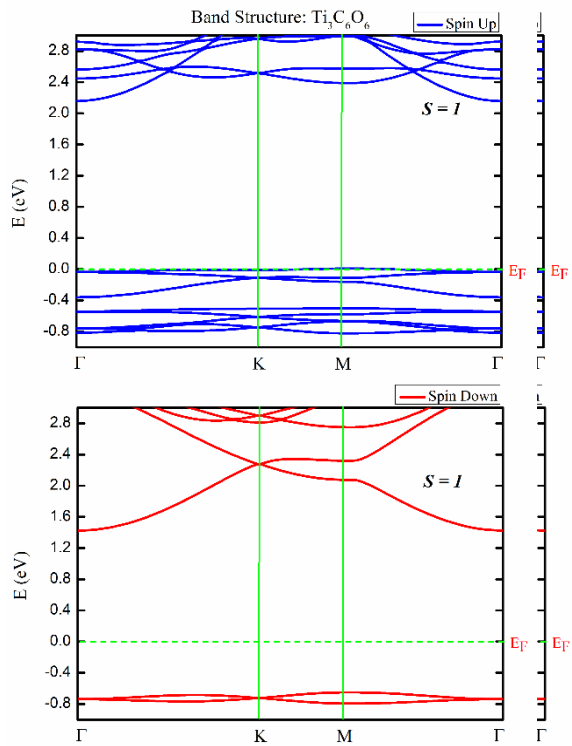
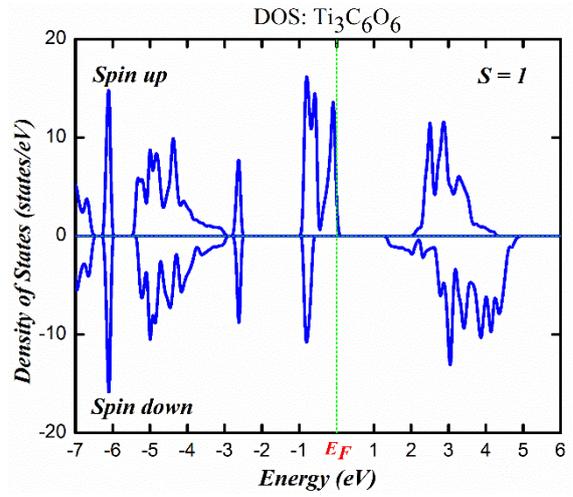
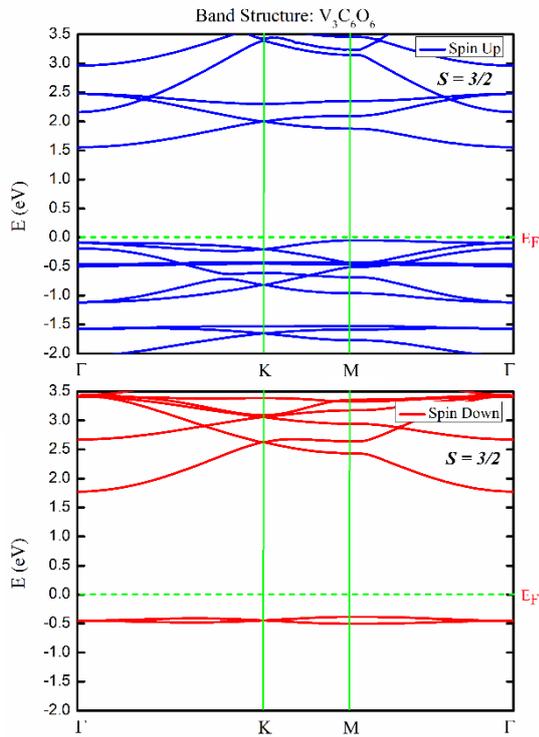
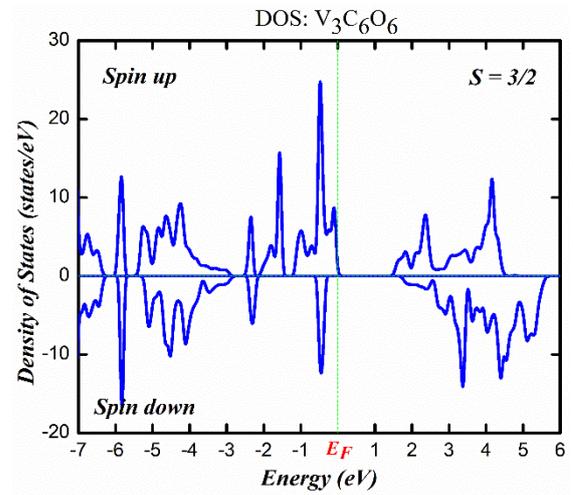

***Figure 5a:*** *Band-structures and total DOS of the 2D transition-metal (TM) - $C_6O_6$ networks with the PAW-SGGA+U method, when TM is Sc, Ti, and V, respectively.*

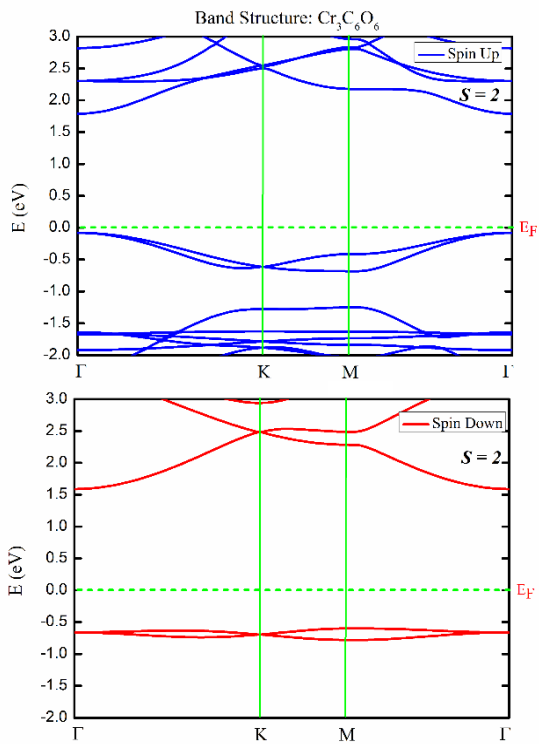
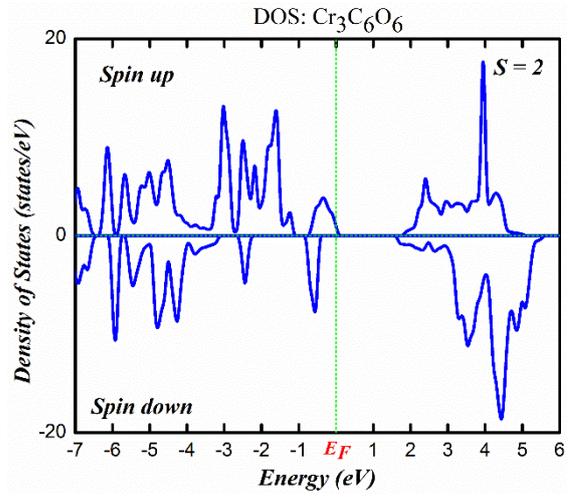
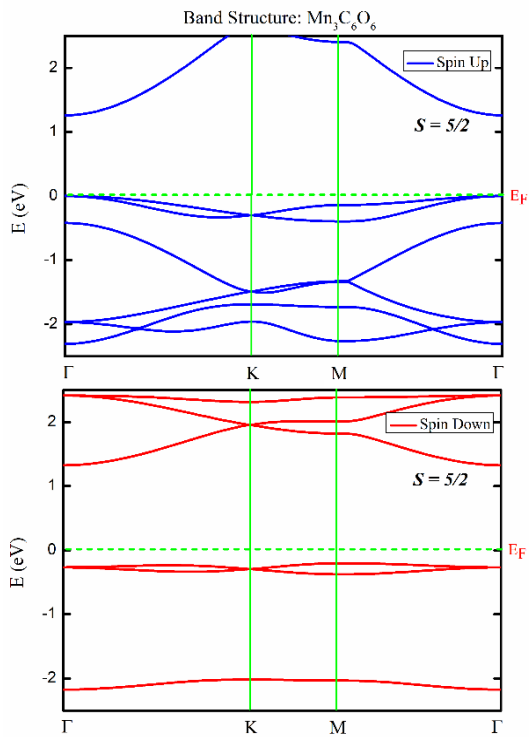
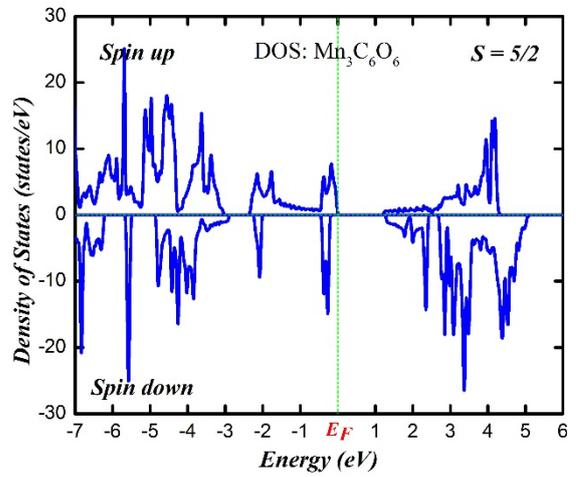

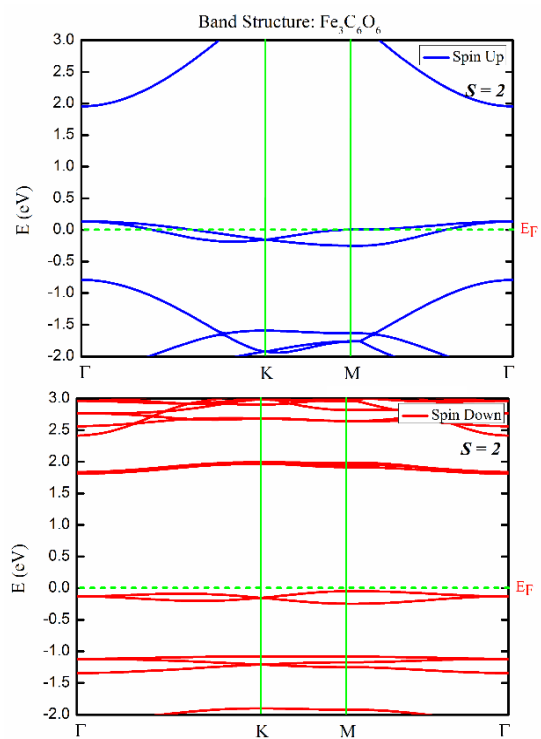 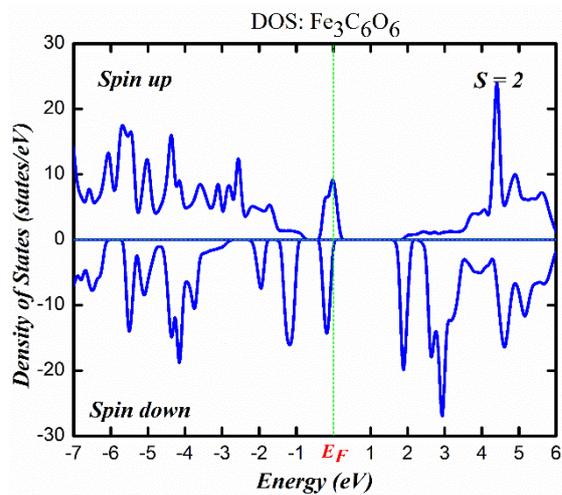

*Figure 5b:* *The same as Fig. 5a but for the TMs Cr, Mn, and Fe.*

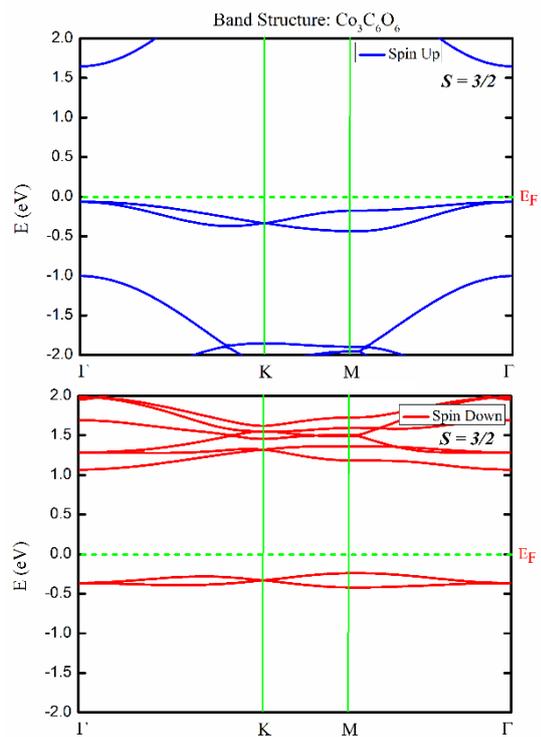 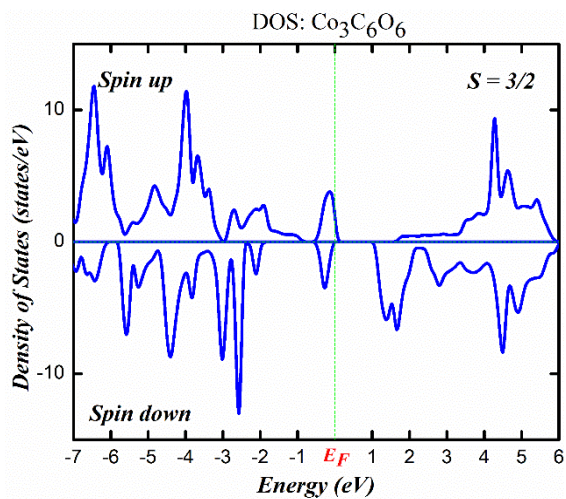

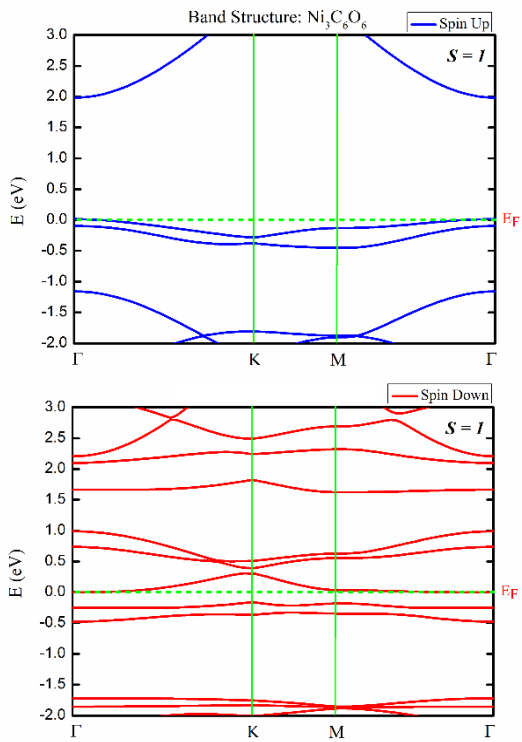
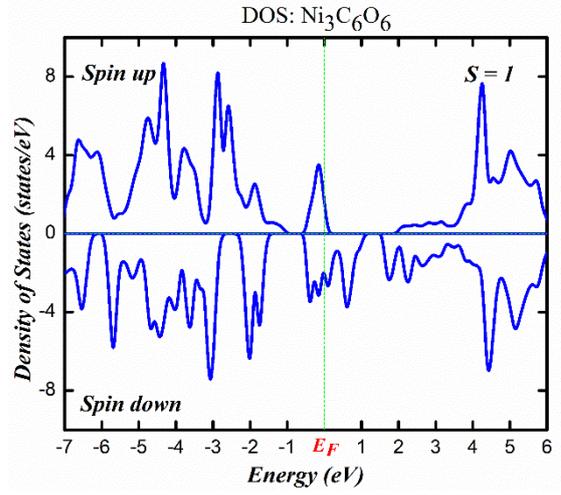
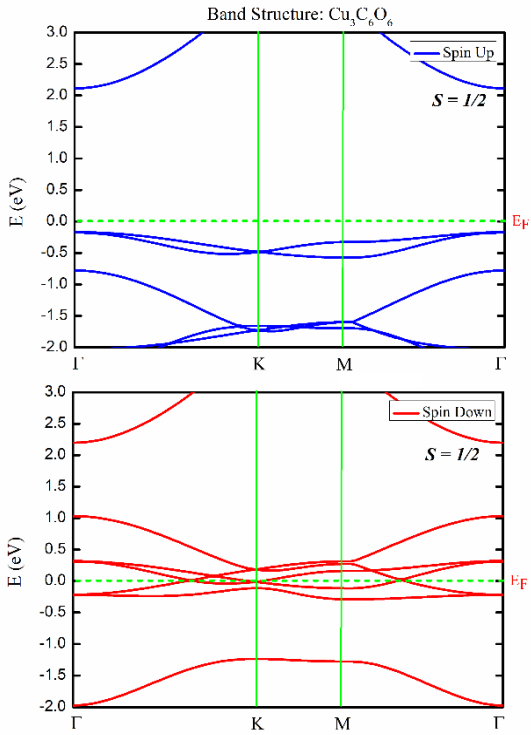
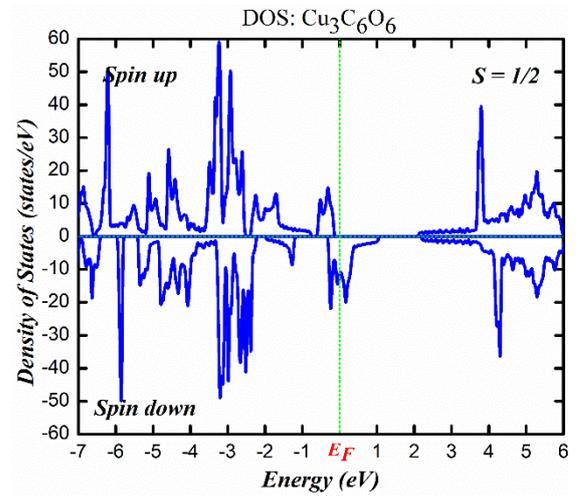

***Figure 5c:*** *The same as Fig. 5a but for the TMs Co, Ni, and Cu.*

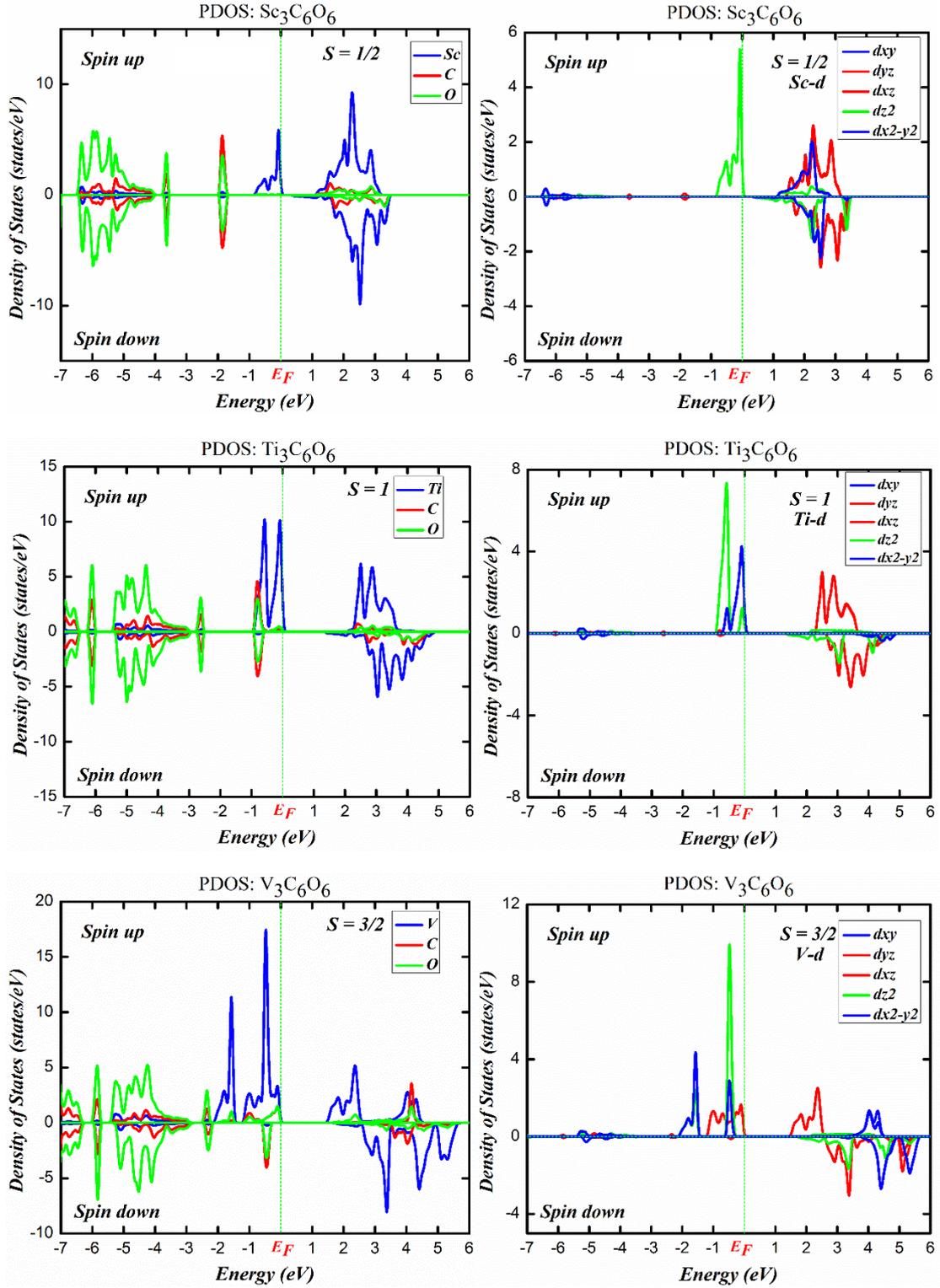

***Figure 6a:*** *Projected DOS of the atoms (TM, C and O) and d orbitals on the TM atoms in the 2D TM$_3$C$_6$O$_6$ materials with the PAW-SGGA+U method for the TMs Sc, Ti, and V.*

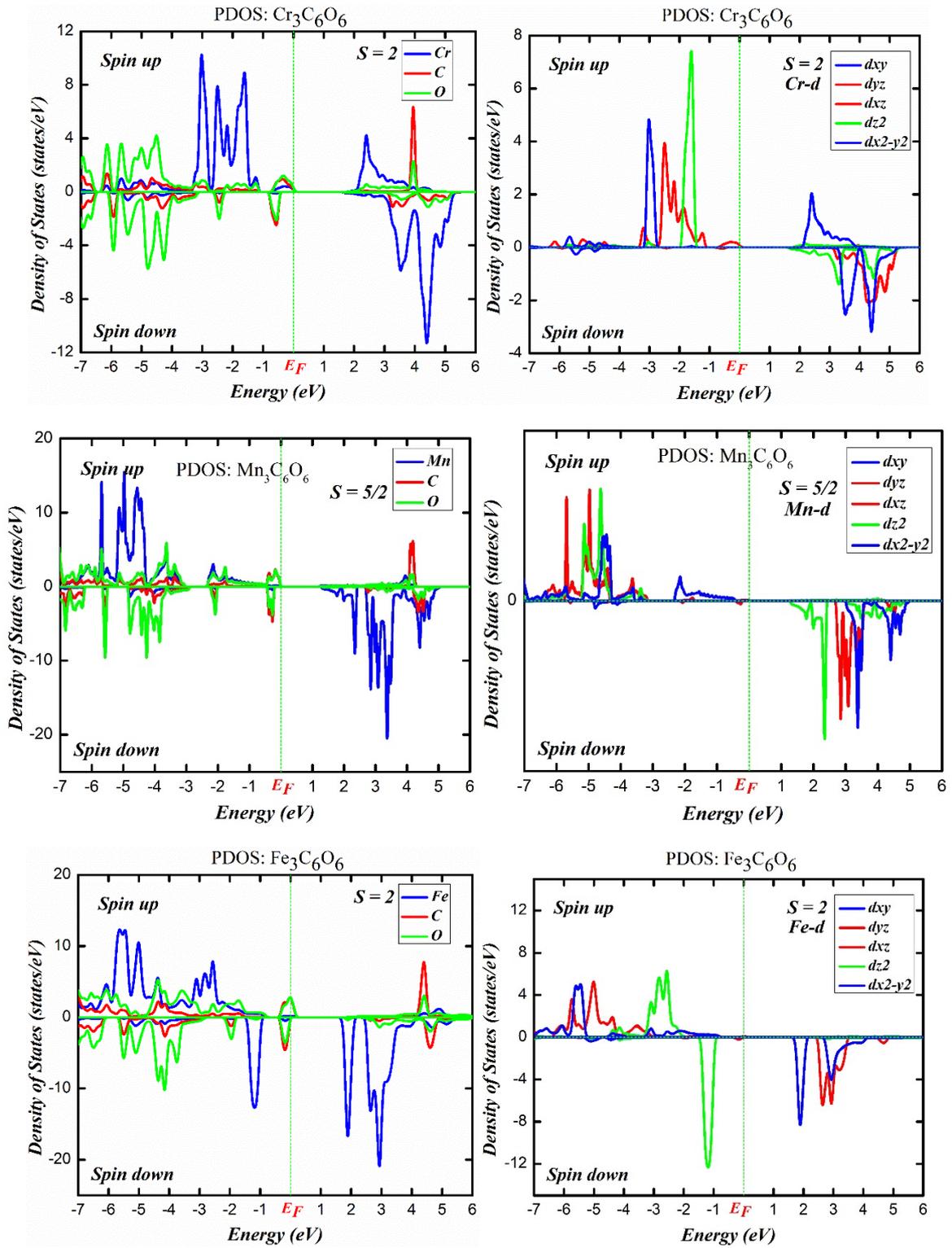

*Figure 6b:* The same as Fig. 6a but for the TMs Cr, Mn and Fe.

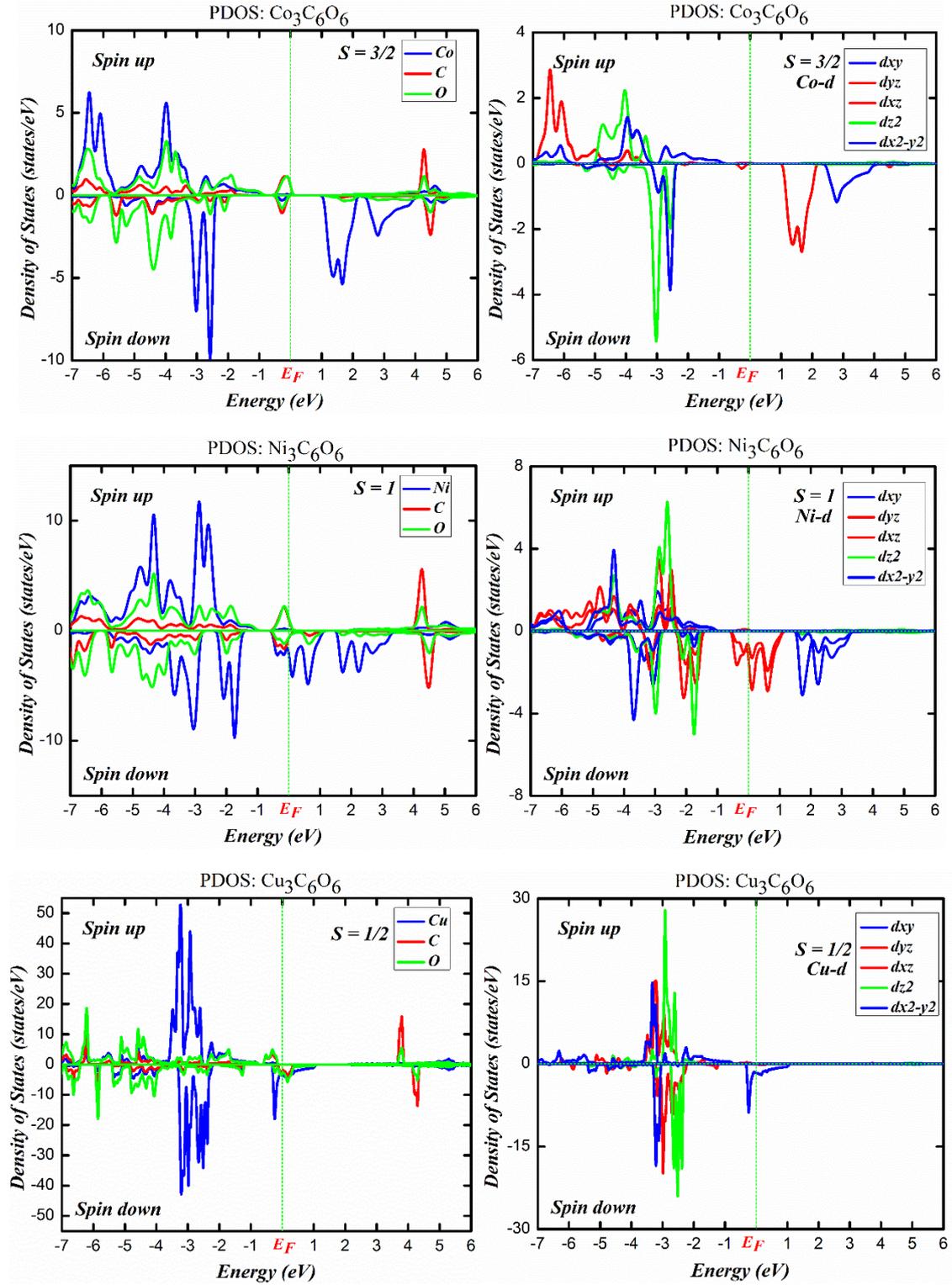

***Figure 6c:*** *The same as Fig. 6a but for the TMs Co, Ni, and Cu.*

The $Cu_3C_6O_6$ case is highly interesting, since it is metallic and shows band crossings at the Fermi level without spin-orbit coupling for parallel spin arrangements. It has ferromagnetic second neighbor couplings which dominate with respect to the nearest neighbor ones and we predict a free-standing $Cu_3C_6O_6$ layer to be a ferromagnetic kagome metal. According to Fig. 5c, it has three band-crossing points at the Fermi level along the path Γ-K-M-Γ. Regarding more in detail (see Appendix B), it becomes clear that the crossing between Γ and K is connected with the crossing between Γ and M, whereas the band-crossing at K is isolated. Finally, we have a Fermi line around Γ and two points at K and K' in the Brillouin zone (shown in Fig. 7a). As expected, the SO coupling with the magnetization perpendicular to the plane opens small gaps of about 10 meV at the crossing points which is shown in Fig. 7b. Similar to the kagome lattice $Cs_2LiMn_3F_{12}$ [9] we expect also $Cu_3C_6O_6$ to be a Chern insulator and to show the QAHE.

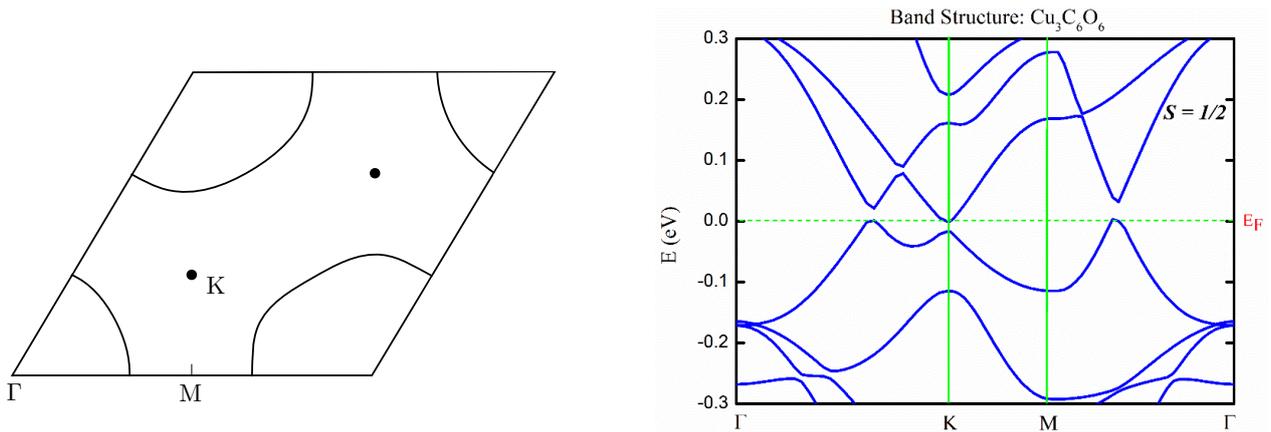

*Figure 7*: *Fermi surface (a, left) without SO coupling consisting of a circle around Γ and two isolated points in the Brillouin zone, and band-structure (b, right) including SO coupling which opens a gap for the $Cu_3C_6O_6$ network calculated with the PAW-SGGA+U method.*

**SUMMARY AND DISCUSSION**

We summarize the possible electronic states and magnetic couplings of the $TM_3C_6O_6$ material class in Tab. 2. Possible spin-crossover (SCO) or magnetic-nonmagnetic transitions for the Fe, Co, Ni, and Cu compounds (see Fig. 1) are not mentioned in this table. Due to the SCO transition the variation of the total energy with the lattice constant is smaller than without it, which should allow epitaxial growth on very different lattices as it was observed for Fe-Zwitterionic Quinone (Fe-ZQ) on Ag(111) and Au(110) surfaces [76]. There is a series of semiconducting compounds with dominating antiferromagnetic nearest neighbor exchange but varying local spins between $S=1/2$ and $S=5/2$, which is very interesting since it could allow experimental studies of the antiferromagnetic kagome lattice for different spins. Three compounds are predicted to be metallic with the interesting perspectives to investigate the interaction between magnetic order and transport properties, especially the anomalous Hall effect. As we argued, the anomalous Hall effect can even be quantified for $Cu_3C_6O_6$. For a final conclusion about the possible magnetic states, also local magnetic anisotropies should be investigated in further studies. As one can see by comparing Tables 1 and A1, the structural properties and the values of the local magnetic moments are much less influenced by the Hubbard $U$ correction than the electronic properties, especially the gap values. Therefore, the present theoretical predictions have to be verified by experimental methods.

*Table 2.* *Summary of electronic and magnetic properties of the $TM_3C_6O_6$ material class. Mentioned are either the dominating magnetic couplings or both of them if they are of comparable strength. The results for $Mn_3C_6O_6$ are taken from* [22]. *Abbreviations that are used in the table: FM - ferromagnetic, AFM - antiferromagnetic, SC - semiconductor, and M -metal.*

| System | Local spin | Magnetic couplings (dominant) | Electronic properties |
|---|---|---|---|
| $Sc_3C_6O_6$ | 1/2 | AFM (1st neighbor) | SC |
| $Ti_3C_6O_6$ | 1 | AFM (1st neighbor) | SC |
| $V_3C_6O_6$ | 3/2 | FM (1st neighbor) | SC |
| $Cr_3C_6O_6$ | 2 | AFM (1st neighbor) | SC |
| $Mn_3C_6O_6$ | 5/2 | AFM (1st neighbor) | SC |
| $Fe_3C_6O_6$ | 2 | AFM (1st neighbor) | M |
| $Co_3C_6O_6$ | 3/2 | FM (1st) and AFM (2nd neighbor) | SC |
| $Ni_3C_6O_6$ | 1 | AFM (1st and 2nd neighbor) | M |
| $Cu_3C_6O_6$ | 1/2 | FM (no Heisenberg) | M |

**CONCLUSION AND PERSPECTIVES**

We investigated systematically with the help of first principles calculations the structural, electronic, and magnetic properties of organometallic networks, which are built of $C_6O_6$ ligands (L) and 3d transition metals (TM) and detected a surprisingly rich behavior. Similar organometallic networks with other organic molecules are also known [85,86]. We investigated the $TM_3L$ material class, for calculations of the $TM_3L_2$ class with a larger distance between TM ions and correspondingly weaker magnetic couplings, see the recent publication [87]. We have found that for $Fe_3C_6O_6$, $Co_3C_6O_6$, $Ni_3C_6O_6$ and $Cu_3C_6O_6$ the magnetic state changes as a function of the lattice parameter. These transitions may be triggered by different substrates or other external stimuli like pressure or temperature.

Three of the calculated monolayers (for the Fe [23], Mn [22], and Cu [21] systems) could already be synthesized on noble metal substrates. There is no visible reason which could hinder to synthesize also all the other investigated materials, and it would be interesting to try insulating substrates. A metallic substrate is expected to influence the electronic structure in several respects: it may fix the lattice constant, induce a charge transfer between substrate and monolayer, and lead to a slight buckling of the monolayer. However, as it was calculated for $Mn_3C_6O_6$ [22], the local magnetic moments are expected to remain with slightly changed magnetic interactions. Investigating the magnetic properties of $Sc_3C_6O_6$ monolayers one could expect signatures of spin liquid behavior since we predict it to be a $S=1/2$ kagome Heisenberg antiferromagnet. Possible experimental methods could be X-ray dichroism, spin resolved tunnelling spectroscopy, or magneto-optical Kerr studies. Interesting is the comparison with $Ti_3C_6O_6$ which is also a kagome antiferromagnet but with $S=1$. The ferromagnetic nearest neighbor couplings in $V_3C_6O_6$ are also interesting. Together with an appropriate magnetic anisotropy (still to be calculated) they could lead to spin ice behavior.

2D metal-organic conducting systems are important for technical applications and we predict metallic behavior for the Fe, Ni and Cu metal-organic compounds. From a fundamental point of view, a band-crossing at the Fermi level allows to study topological quantum states, that we found to be realized in $Cu_3C_6O_6$ (without SO coupling) in connection with ferromagnetic exchange couplings. The SO coupling leads to small gaps of about 10 meV and this material is a candidate for a Chern insulating phase and the QAHE. Comparing the free-standing layer of $Cu_3C_6O_6$ investigated here with the one synthesized on Ag(111) [22], there is unfortunately a charge transfer for the adsorbed layer that destabilizes the magnetic solution with respect to the nonmagnetic one with larger lattice constant. So, it would be interesting to look for alternative synthesis routes of this promising material using for instance insulating substrates.

## AUTHOR INFORMATION

**Corresponding Author**

*E-mail: hassan.denawi@cea.fr (H.D.)

## ACKNOWLEDGEMENTS

This work was supported by the computer resources of the Centre Informatique National de l'Enseignement Supérieur (CINES), Project No. A0020906873 and the High-Performance Computing (HPC) resources of Aix-Marseille University financed by the project Equip@Meso (ANR-10-EQPX-29-01). HD and XB mention that this study has been (partially) supported through the EUR grant NanoX n°ANR-17-EURE-0009 in the framework of the "Programme des Investissements d'Avenir". We thank M. Koudia, V. Gubanov, O. Janson, I. Makhfudz and O. Siri for helpful discussions.

## Conflict of interest

The authors declare that they have no conflict of interest.

# Appendix A

The ab-initio results without Hubbard $U$ correction are listed in Table A1.

**Table A1.** *The same physical parameters as in Table 1 but without U correction.*

| | $d^1$ | $d^2$ | $d^3$ | $d^4$ | $d^5$ | $d^6$ | $d^7$ | $d^8$ | $d^9$ |
|---|---|---|---|---|---|---|---|---|---|
| Without U | $Sc_3C_6O_6$ (S=1/2) | $Ti_3C_6O_6$ (S=1) | $V_3C_6O_6$ (S=3/2) | $Cr_3C_6O_6$ (S=2) | $Mn_3C_6O_6$ (S=5/2) | $Fe_3C_6O_6$ (S=2) | $Co_3C_6O_6$ (S=3/2) | $Ni_3C_6O_6$ (S=1) | $Cu_3C_6O_6$ (S=1/2) |
| $d_{TM-O}$ | 2.14 | 2.09 | 2.04 | 2.03 | 2.10 | 2.04 | 2.02 | 2..03 | 2.04 |
| $a$ (Å) | 8.06 | 7.92 | 7.80 | 7.76 | 7.94 | 7.80 | 7.74 | 7.70 | 7.78 |
| $E_{ex-1}$ | -24.77 | -7.49 | 39.96 | -8.37 | -26.07 | -52.96 | -59.65 | -33.60 | 3.92 |
| $E_{ex-2}$ | -44.47 | -12.83 | 110.95 | -14.97 | -46.35 | -91.60 | 31.14 | -39.32 | 3.1 |
| $J_1$ | 59.11 | 4.01 | -23.66 | 1.24 | 2.43 | 7.25 | -30.26 | 4.29 | 2.46 |
| $J_2$ | 7.61 | 0.81 | 5.17 | 0.17 | 0.35 | 1.34 | 25.07 | 10.46 | -7.11 |
| $M$ | 1 | 2 | 3 | 4 | 5 | 4 | 3 | 2 | 1 |
| $M_d$ | 0.404 | 1.337 | 2.30 | 3.27 | 4.36 | 3.49 | 2.44 | 1.36 | 0.378 |
| $M_m$ | 0.442 | 1.410 | 2.33 | 3.31 | 4.45 | 3.52 | 2.46 | 1.37 | 0.376 |
| $E_a$ | 1.21 | 0.83 | 0 | 0.75 | 0.64 | 1.76 | 0 | 0 | 1.99 |
| $E_b$ | 2.21 | 2.20 | 2.37 | 1.95 | 1.08 | 0 | 0 | 0 | 0 |
| $E_g$ | 1.05 | 0.82 | 0 | 0.75 | 0 | 0 | 0 | 0 | 0 |

## Appendix B

To clarify the Fermi surface of $Cu_3C_6O_6$ we calculated a series of band structures along the lines S1 ... S6 starting at $\Gamma$ and ending at $\vec{b_1} + x\vec{b_2}$ with $x = \{0\,;\,0.1\,;\,0.2\,,0.3\,,0.4\,,0.5\}$ (Figure B1). Please remind that $(1/2)\vec{b_1}$ and $(1/2)\vec{b_2}$ correspond to the M point and $(2/3)\vec{b_1} + (1/3)\vec{b_2}$ corresponds to the K point in the Brillouin zone.

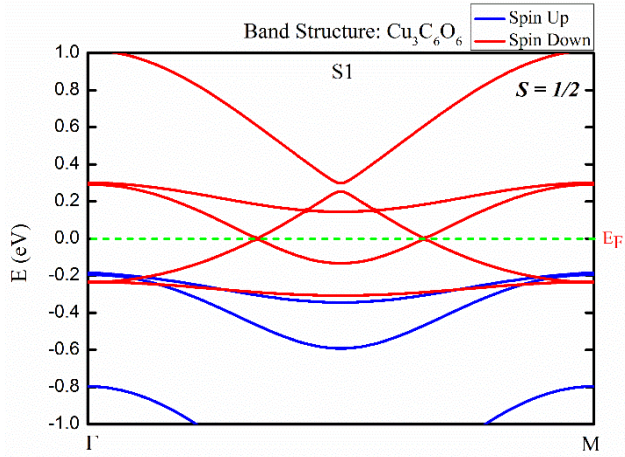
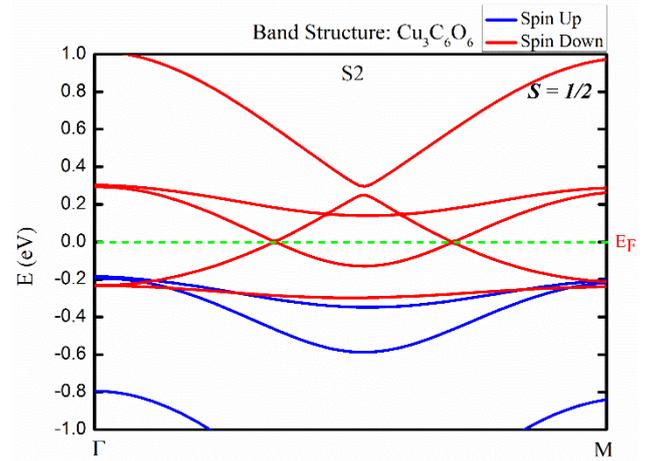
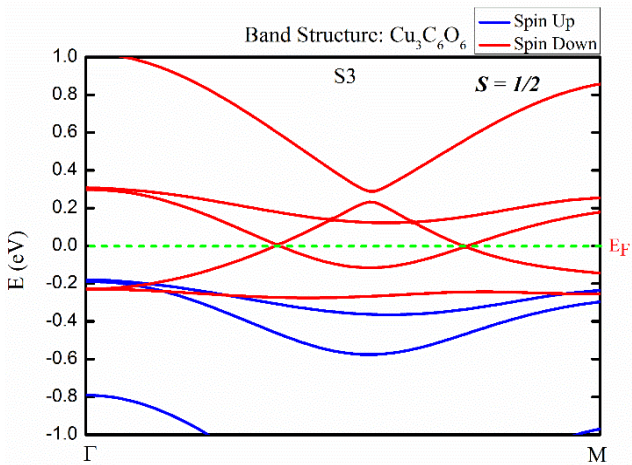
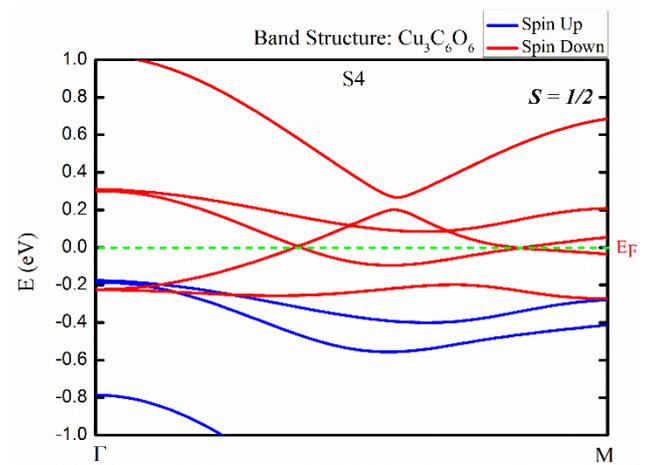

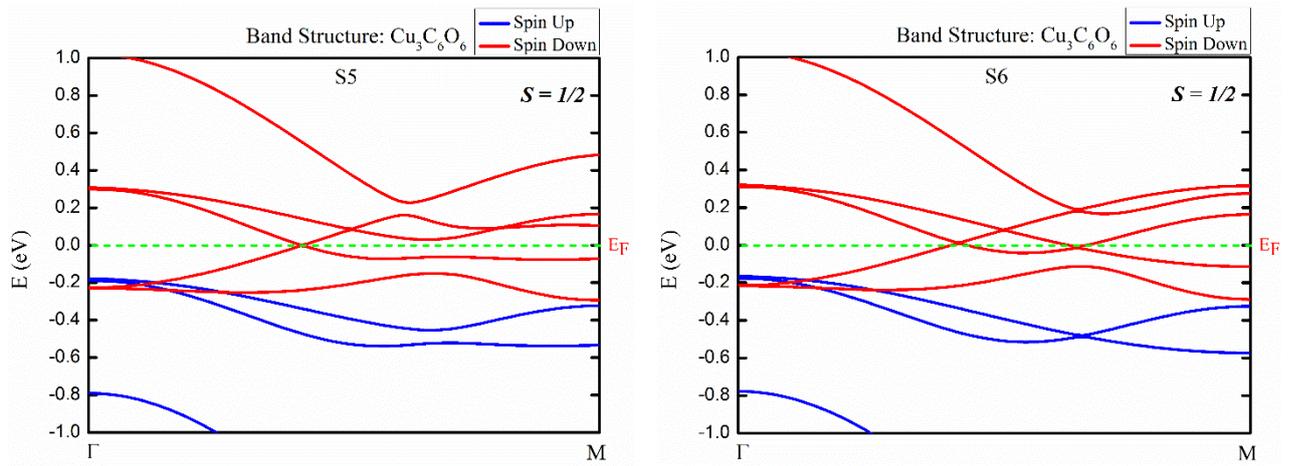

**Figure B1:** Band structure without SO coupling of Cu$_3$C$_6$O$_6$ along the lines S1 ... S6 explained in the text.